\documentclass[prb,showpacs,showkeys,preprintnumbers,amsmath,amssymb,twocolumn]{revtex4-1}
\usepackage[T1]{fontenc} 
\usepackage{makeidx} 
\usepackage{graphicx} 
\usepackage{dcolumn} 
\usepackage{array} 
\usepackage{amssymb} 
\usepackage{amsmath}
\usepackage{textcomp}
\usepackage{rotating}
\usepackage{wasysym}
\usepackage{multirow}
\usepackage{subfigure}
\usepackage{color}
\usepackage{eucal}
\usepackage{mathrsfs}
\usepackage{units}
\usepackage{rotating}
\usepackage[all]{xy}
\usepackage{float} 
\usepackage{amsmath} 
\usepackage{amsfonts} 
\usepackage{bm} 
\usepackage[amssymb]{SIunits}
\definecolor{ashgrey}{rgb}{0.7, 0.75, 0.71}

\begin{document}

\title{Quantum capacitance oscillations in graphene under crossed magnetic and electric fields}

\author{Z.Z. Alisultanov}\email{zaur0102@gmail.com}
\affiliation{Amirkhanov Institute of Physics Russian Academy of Sciences, Dagestan Science Centre, Russia, 367003, Makhachkala, Yagarskogo str., 94}
\affiliation{Prokhorov General Physics Institute Russian Academy of Sciences, Russia, 119991, Moscow, Vavilov Str., 38}
\affiliation{Dagestan States University, Russia, 367000, Makhachkala, Gadzhiyev Str., 43-a}
\author{M.S. Reis}\email{marior@if.uff.br}\affiliation{Instituto de F\'{i}sica, Universidade Federal Fluminense, Av. Gal. Milton Tavares de Souza s/n, 24210-346, Niter\'{o}i-RJ, Brasil}

\date{\today}

\begin{abstract}
Quantum oscillations of metallic systems at low temperatures is one of the key rules to experimentally access their electronic properties, such as energy spectrum, scattering mechanisms, geometry of Fermi surface and many other features. The importance of these knowledge is enormous, since from these a thorough understanding of anomalous Hall effect, thermopower and Nernst coefficients, just to name a few, is possible; and from those knowledge, a plenty of applications arise as emerging technologies. In this direction, the present contribution focus on a complete description of quantum capacitance oscillations of monolayer and bilayer graphenes under crossed electric and magnetic fields. We found a closed theoretical expression for the quantum capacitance and highlight their amplitude, period and phase - important parameters to access the electronic properties of graphenes. These results open doors for further experimental studies.
\end{abstract}

\maketitle

Magnetic and conductive thermal properties of graphenes are some of the actual problems on experimental and theoretical condensed matter physics\cite{r1}; mainly due to its two-dimensional lattice and Dirac spectrum of energy, leading thus to uniques features. The magnetic field is responsible to the unusual energy spectrum of graphenes\cite{r2}, since its Landau levels (LLs) are nonequidistant (the gap between the first two LLs in a magnetic field of 10 T is larger than 1000 K), and then remarkable effects arise, such as the giant magneto \cite{r3,r4} and thermo-magnetic effects \cite{r5}, as well as the unusual quantum Hall effect, which can be observed even at room temperature \cite{r1,r6,r7}; and these facts make therefore graphenes a promising material for modern nanoelectronics. The oscillating magnetocaloric effect is also remarkable, since the normal and inverse effects can be tuned by the final value of applied magnetic field \cite{cinco,alisultanov2015oscillating,alisultanov2014oscillating,alisultanov2015oscillating,paixao2014oscillating,reis2014diamagnetic,reis2014step,reis2013oscillating,reis2013electrocaloric,reis2013influence,reis2012oscillating,reis2012oscillating,reis2011oscillating}; and, from this normal-to-inverse effect, an enhanced  thermomagnetic hexacycle was proposed\cite{reis2015magnetocaloric}.

One key rule of the properties above described (and even many others), is the density of states (DOS), that is linearly dependent on the energy; and for undoped graphenes, the DOS vanishes at the Fermi level (corresponding thus to the Dirac point). A thorough experimental and theoretical study of this feature shall then be absolutely necessary, since from those findings, deeper and advanced technological achievements are possible. In this direction, one experimental method to study the DOS is based on the measurement of quantum oscillations of physical parameters in high magnetic field, to then access their electronic spectrum, scattering mechanisms, geometry of Fermi surface and many other information.

Based on above description, the present Letter focus to investigate the oscillatory properties of quantum capacitance of graphenes under crossed magnetic and electric fields. Similar studies have been carried out to explore magnetization\cite{r16}  and electrocaloric effect\cite{reis2013electrocaloric}; and all these are indeed useful for further experimental studies and then deepen our knowledge on the DOS of graphenes. All of these are important challenges to overcome for a thorough understanding of the behavior of these materials and plan new applications.

For the present Letter we considered (i) the quasi-classical approach based on the quantization conditions of Lifshitz-Onsager\cite{r15,r18}, (ii) the graphene sheet on the $xy$ plane, (iii) the magnetic field as $\vec{H}=(0,0,H)$ and (iv) the electric field as $\vec{E}=(E,0,0)$. From these assumptions, the quantized area enclosed by an electron trajectory in momentum space is given by\cite{r19,r20}:
\begin{equation}\label{A1}
A(\epsilon^\prime)=\frac{2\pi \hbar eH}{c}(j+\gamma_\pm)
\end{equation}
where $j$ is an integer, $e$ represents the electron charge, $c=3\times10^{10}$ cm/s the light speed and
\begin{equation}
\gamma_\pm=\gamma\pm\frac{m^*}{2m}.
\end{equation}
Above, 
\begin{equation}\label{mcyy}
m^*=m(\epsilon^\prime)=\frac{1}{2\pi}\frac{dA(\epsilon^\prime)}{d\epsilon^\prime},
\end{equation}
is the electron cyclotron mass; $\gamma$ is a constant that defines the energy of the zeroth Landau level (it assumes $1/2$ for non-relativistic gas and zero for graphenes); $m^*/2m$ is the Zeeman splitting and $m$ is the electron mass. Similarly to what was done before\cite{alisultanov2015magneto}, in the present work we also ignored the Zeeman splitting of the Landau levels.  Moreover, 
\begin{equation}\label{ene2}
\epsilon^\prime=\epsilon-v_0p_y
\end{equation}
where $\vec{v_0}=cE\hat{y}/H$ is the average drift velocity of the electron perpendicularly to $EH$ plane and 
\begin{equation}\label{eMG}
\epsilon=\nu_bv_Fp
\end{equation} 
is the energy spectrum of a monolayer graphene MG in the vicinity of Dirac point. In addition, $v_F\approx 10^8$ cm/s is the Fermi velocity and $\nu_b=\pm1$ is the band index (`+' for the conduction band and `-' for the valence band). 

From the above, it is easy to show that a curve with constant energy $\epsilon^\prime$ is an ellipse on the $p_xp_y$ plane, with the following parameters: $a=\epsilon^\prime/\sqrt{v_F^2-v_0^2}$ and $b=\epsilon^\prime v_F/(v_F^2-v_0^2)$. Then the quantized area on the momentum space reads as:
\begin{equation}\label{A2}
A(\epsilon^\prime)=\pi ab=\frac{\pi\epsilon^{\prime2}v_F}{(v_F^2-v_0^2)^{3/2}}
\end{equation}
From a simple comparison of equations \ref{A1} and \ref{A2} (and the help of equation \ref{ene2}), we obtain the final energy spectra of the graphene sheet under consideration (i.e., on the $xy$ plane under $\vec{H}=(0,0,H)$ and $\vec{E}=(E,0,0)$):
\begin{equation}
\epsilon_{j,p_y}=\text{sgn}(j)(1-\beta^2)^{3/4}\frac{\hbar v_F}{l_H}\sqrt{2|j|}+v_0p_y
\end{equation}
where $\beta=v_0/v_F$ and $l_H=\sqrt{\hbar c/eH}$ is the characteristic magnetic length. The expression above completely coincides with the expression obtained by solving the Dirac equation\cite{r13}. 

To the sake of comparison, a bilayer graphene BG is also considered. To this purpose, let us follow the same procedure as before. Thus, to go further,  we must recall the energy spectrum of a BG in the vicinity of Dirac point (analogously to equation \ref{eMG})\cite{r4}:
\begin{equation}
\epsilon=\nu_b\left[\sqrt{\left(v_F^2p^2+\frac{t^2_\perp}{4}\right)}-\frac{\nu_{sb}t_\perp}{2}\right]
\end{equation}
where $t_\perp$ is the effective hopping energy between two layers (analogously to our recent contribution\cite{alisultanov2015oscillating}, we considered $t_\perp$ close to 0.4 eV.), and $\nu_{sb}=\pm1$  is the bilayer sub-band index.  As before, considering again a constant value of $\epsilon^\prime$, the quantized area on the momentum space reads now as:
\begin{equation}\label{A3}
A(\epsilon^\prime)=\pi\frac{\epsilon^{\prime2}+\epsilon^\prime t_\perp+\beta^2t_\perp^2/4}{v_F^2(1-\beta^2)^{3/2}}
\end{equation}
Again, a simple comparison of equations \ref{A1} and \ref{A3} (and the help of equation \ref{ene2}), lead us to the final energy spectra of the BG under consideration (i.e., on the $xy$ plane under $\vec{H}=(0,0,H)$ and $\vec{E}=(E,0,0)$):
\begin{equation}
\epsilon_{j,p_y}=\text{sgn}(j)\sqrt{(1-\beta^2)^{3/2}\frac{\hbar^2v_F^2}{l_H^2}2|j|+\frac{t_\perp^2}{4}(1-\beta^2)}+v_0p_y
\end{equation}


From those energy spectra, we are now able to focus on our aim: \textit{quantum capacitance}, that is defined as\cite{r10,r11,r12}:
\begin{equation}\label{definicaoC}
C=-e^2\int_0^\infty\frac{\partial f}{\partial\epsilon}\rho(\epsilon)d\epsilon
\end{equation}
where
\begin{equation}
f(\epsilon)=\frac{1}{\exp[(\epsilon-\mu)/k_BT]+1}
\end{equation}
is the Fermi-Dirac distribution function,  $\mu$ is the chemical potential and
\begin{equation}
\rho(\epsilon)=-\frac{1}{\pi}\text{sgn}(\epsilon-\mu)\text{Im}\{G(\epsilon)\}
\end{equation}
the density of states. Above, $G(\epsilon)$ is the single-particle Green function
\begin{equation}
G(\epsilon)=\sum_{j,p_y}\frac{1}{\epsilon-\epsilon_{j,p_y}+i\;\text{sgn}(\epsilon-\mu)\Gamma}
\end{equation}
where $\Gamma$ is the width of the LLs related to the scattering on impurities. Thus, a minor calculation leads to:
\begin{equation}
\rho(\epsilon)=\frac{1}{\pi}\sum_{j,p_y}\frac{\Gamma}{(\epsilon-\epsilon_{j,p_y})^2+\Gamma^2}
\end{equation}

To go further, we have used the Poisson summation formula and have obtained:
\begin{equation}
\rho(\epsilon)=\rho_0(\epsilon)+\rho_{osc}(\epsilon)
\end{equation}
where
\begin{equation}
\rho_0(\epsilon)=\frac{L_y}{\pi^2\hbar}\int_0^{p_{ymax}}dp_y\int_0^\infty\frac{\Gamma}{(\epsilon-\epsilon_{x,p_y})^2+\Gamma^2}dx
\end{equation}
and
\begin{equation}\label{osciii}
\rho_{osc}(\epsilon)=\frac{2L_y}{\pi^2\hbar}\text{Re}\left\{\sum_{k=1}^\infty\int_0^{p_{ymax}}dp_y\int_0^\infty\frac{\Gamma e^{-i2\pi kx}}{(\epsilon-\epsilon_{x,p_y})^2+\Gamma^2}dx\right\}
\end{equation}
where $L_y$  is the graphene size along $y$ axis and $p_{ymax}$ is determined from the condition of the degeneracy of the LLs\cite{r13}:
\begin{equation}
0< x=\frac{c}{eH}p_y<L_x
\end{equation}
From the above, it is easy to see that $p_{ymax}=eHL_x/c$. 


Some assumptions now shall be done: $\mu\gg k_BT$ and $\Gamma\rightarrow0$. The first one makes the integrand to be significantly different from zero only near the point $\epsilon=\mu$; and therefore only energies $\epsilon\approx\mu$ are important for the magnetic oscillations. From these assumptions and after some evaluations (with the help of equations \ref{A1} and \ref{mcyy}), we obtain:
\begin{equation}
\rho_0(\epsilon)=\frac{cL_y}{2\pi^2\hbar^2 eH}\int_0^{eHL_x/c}dp_ym^*_0\Theta(\epsilon-v_0p_y-\epsilon_0)
\end{equation}
where $m^*_0=m(\epsilon-v_0p_y)$ and $\epsilon_0=\epsilon_{j=0}$. Considering also $\mu\gg eEL_x$, the zero-field contribution to the density of states reads then as:
\begin{equation}\label{zerofinal}
\rho_0(\epsilon)\approx\frac{L_xL_y}{\pi\hbar^2}m^*
\end{equation}

Let us now focus our attention to the oscillating term of the density of states (equation \ref{osciii}). Considering again equations \ref{A1} and \ref{mcyy}, as well as moving the integration over the energy variable, we obtain:
\begin{widetext}
\begin{equation}\label{grands}
\rho_{osc}(\epsilon)=\frac{2L_y}{\pi\hbar^2}\frac{c}{eH}\text{Re}\left\{\sum_{k=1}^\infty\int_0^{p_{ymax}}dp_y\int_0^\infty\delta(\epsilon-\epsilon_{x,p_y})
\exp\left[-ik\frac{c}{\hbar eH}A(\epsilon_x)\right]m(\epsilon_x)d\epsilon_x\right\}
\end{equation}
\end{widetext}
As mentioned above, the relevant values are those that satisfy $\epsilon\approx\mu$; and therefore we expand the function $A(\epsilon)$ in the vicinity of $\mu$:
\begin{equation}
A(\epsilon)\approx A(\mu)+2\pi m^*(\epsilon-\mu)
\end{equation}
After some calculations and considering the above equation we obtain:
\begin{widetext}
\begin{equation}
\rho_{osc}(\epsilon)=\frac{2L_y}{\pi^2\hbar v_0}\sum_{k=1}^\infty\frac{1}{k}\sin\left(k\frac{\pi L_xm(v_0,\mu)v_0}{\hbar}\right)
\cos\left\{\frac{kc}{\hbar e H}\left[A(v_0,\mu)+2\pi m(v_0,\mu)\left(\epsilon-\mu-\frac{v_0p_{ymax}}{2}\right)\right]\right\}
\end{equation}
\end{widetext}
Previously, we have considered $\mu\gg eEL_x=L_xv_0\hbar/l_H^2$ and therefore  the above equation can be rewritten as:
\begin{equation}\label{oscfinal}
\rho_{osc}(\epsilon)=\frac{2L_xL_ym^*}{\pi\hbar^2}\sum_{k=1}^\infty\cos\left\{\frac{kc}{\hbar eH}[A(\mu)+2\pi m^*(\epsilon-\mu)]\right\}
\end{equation}

From equations \ref{definicaoC}, \ref{zerofinal} and \ref{oscfinal}, the quantum capacitance can be written as: 
\begin{equation}\label{qc}
C=\frac{e^2L_xL_ym^*}{\pi\hbar^2}\left\{1+4\sum_{k=1}^\infty
x_k\frac{\cos\left[\frac{kc}{\hbar eH}A(\mu)\right]}{\sinh x_k}\right\}
\end{equation}
where
\begin{equation}
x_k=k\frac{4\pi^2m^*ck_BT}{\hbar eH}
\end{equation}

\begin{figure}
\begin{center}
\subfigure{\includegraphics[width=7cm]{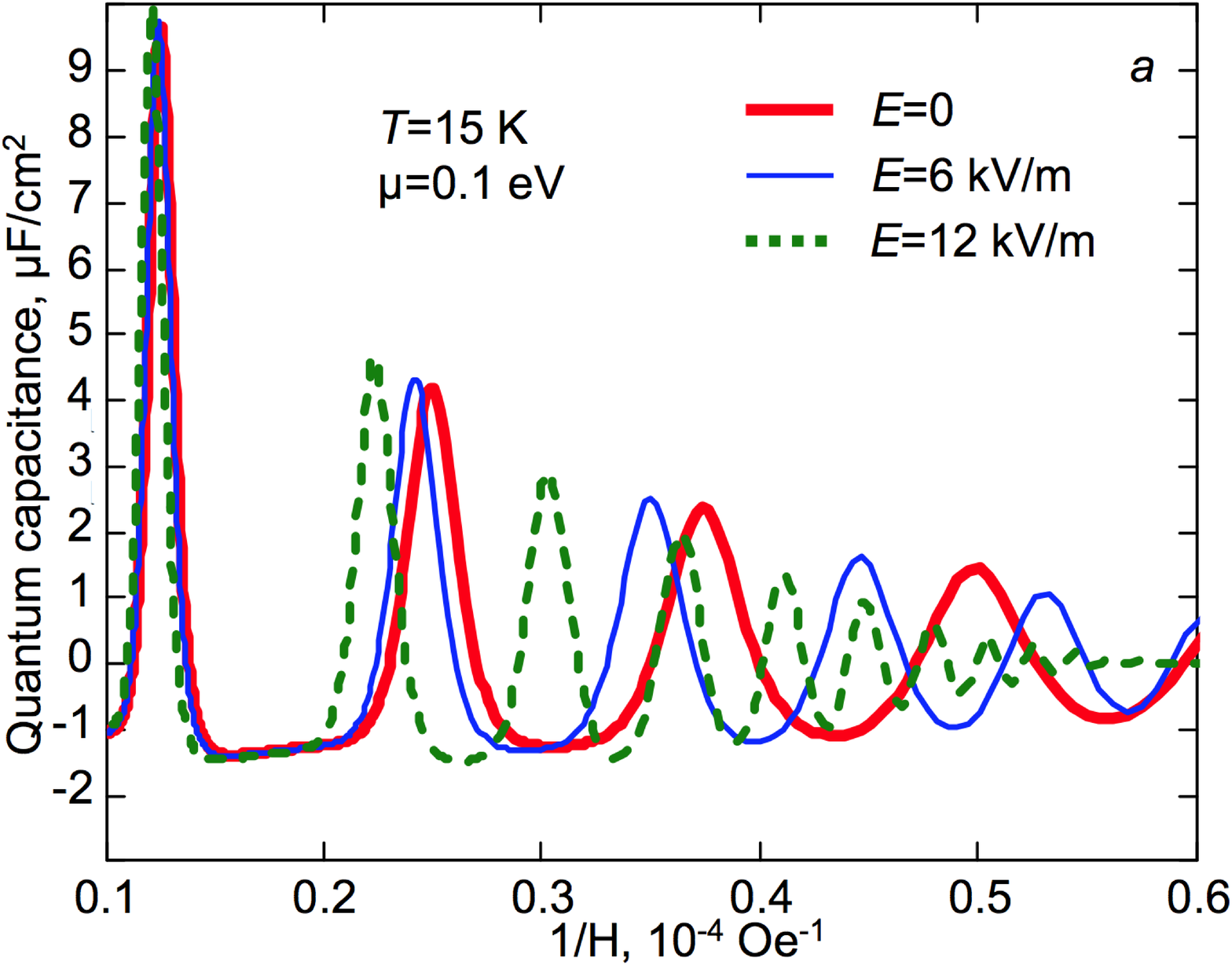}}
\subfigure{\includegraphics[width=7cm]{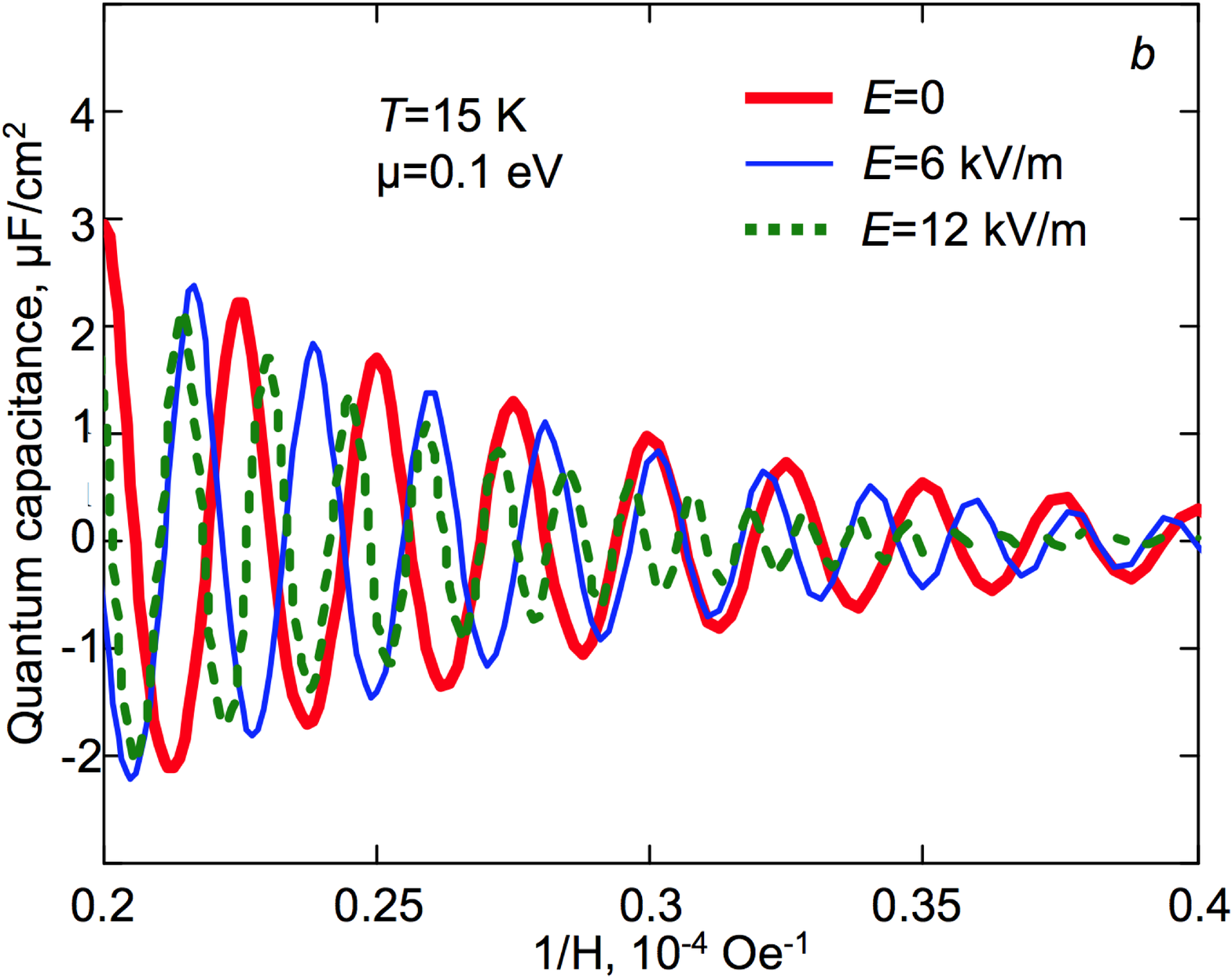}}
\end{center}
\caption{Quantum capacitance for (a) MG and (b) BG, both as a function of reciprocal magnetic field $1/H$. Some values of applied electrical field $E$ is considered, as well as the temperature (15 K) and the chemical potential (0.1 eV, corresponding to 1200 Kk$_\text{B}$). Values of electric field $E$ were chosen in such a way to satisfy $\beta<1$.\label{umfig}}
\end{figure}

From the above result, it is easy to see that quantum capacitance oscillates as a function of the reciprocal magnetic field $1/H$; and the period of these oscillations depend on $A(\mu)$, that, on its turn, depends if we consider either a MG or a BG. In this sense, quantum capacitance is depicted on figure \ref{umfig}, for (a) MG and (b) BG. This figure considers some values of applied electrical field $E$; and it is possible to see that $E$ is able to tune the period of those oscillations. In addition, BG graphene oscillates faster then the MG, mainly due to the hopping term.

\begin{figure}
\begin{center}
\includegraphics[width=7cm]{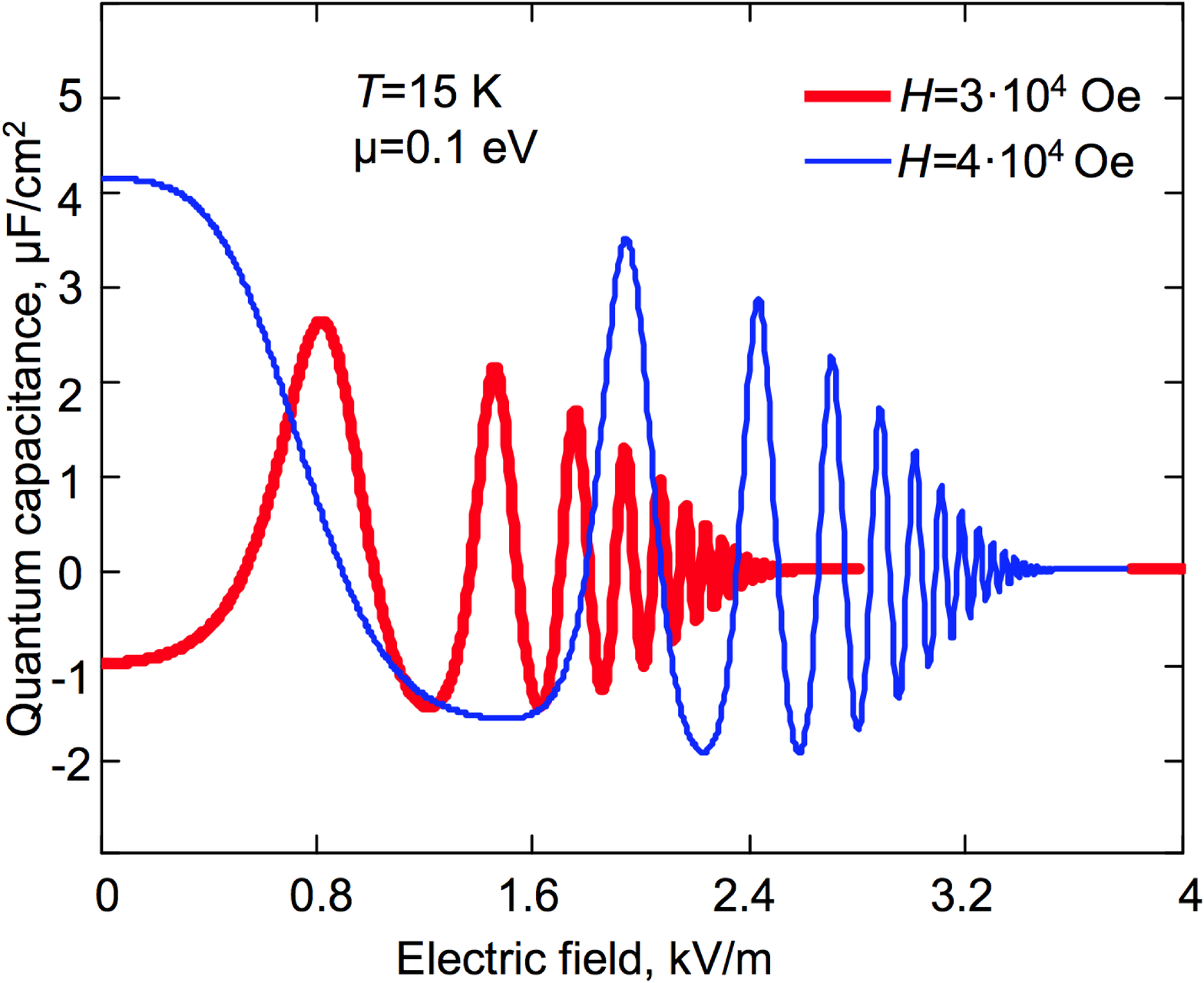}
\end{center}
\caption{Quantum capacitance for a MG as a function of electric field. The aim of this figure is to stress the dependence of the quantum oscillations on the electric field. Parameters used are the same as those of figure \ref{umfig}. \label{doisfig}}
\end{figure}

Considering that the argument of the cosine function on equation \ref{qc} depends on $A(\mu)$; and this one, on its turn, depends also on the electrical applied field, it is natural to see oscillation as a function of $E$, as can be seen in figure \ref{doisfig}. Some values of applied magnetic field is also presented and, in a similar fashion as discussed previously on figure \ref{umfig}, this quantity (magnetic field), also rules the period of oscillations.

A remarkable point shall be addressed, as already discussed in reference \onlinecite{reis2013electrocaloric}, and also found here for the quantum capacitance. If $\beta<1$ we then have all of the oscillations above described; however, if $\beta>1$, the Landau structure collapses and the electronic motion becomes quasi-continuos. The consequence of this last cases is the disappearance of the oscillations. In addition, the influence of the electric field on the Landau structure of graphenes is much more remarkable than in other materials, like semi-conductors\cite{aronov1966indirect}. Thus, as a consequence, tune of the magneto-electronic properties for device applications is indeed much more real for graphenes.

Summarizing, quantum oscillations are a powerful tool to experimentally and theoretically study electronic properties of materials, namely energy spectra, scattering mechanisms, geometry of Fermi surface and much more information; and a deeper knowledge on these quantities gives to the scientific community a better understanding on some anomalous behavior, such as Hall resistivity, thermopower, Nernst coefficient and much more. To crow the importance of these knowledge, these are the basis of several new applications of emerging functional materials, like graphene. In this direction, the present Letter focus on a thorough description of the quantum capacitance oscillation on monolayer and bilayer graphenes under a crossed electric and magnetic fields. We found a closed expression for quantum capacitance of MG and BG and from these results important parameters, like amplitude, period and phase are highlighted as a function of electric and magnetic fields. We are convinced these results are important to guide further experimental studies. Finally, our next step forward is to consider the integer quantum Hall effect, taking into account the dependence of the Landau levels on electric field; considering that this field is able to rebuilt areas of localized states of electrons between these Landau levels.

ZZA declares that this work was supported by grants numbers: RFBR 15-02-03311a, PG MK-4471.2015.2, RSCF MKH-15-19-10049 and 3.1262.2014 from the Ministry of Education and Science of Russia. ZZA is also sincerely grateful to Dmitry Zimin Foundation "Dynasty" for financial support.  MSR thanks FAPERJ, CAPES, CNPq and PROPPI-UFF for financial support. The authors have a deep debit with Prof. RP Meilanov (in memorian) due to his always helpful discussion on the theme.


\begin{thebibliography}{30}%
\makeatletter
\providecommand \@ifxundefined [1]{%
 \@ifx{#1\undefined}
}%
\providecommand \@ifnum [1]{%
 \ifnum #1\expandafter \@firstoftwo
 \else \expandafter \@secondoftwo
 \fi
}%
\providecommand \@ifx [1]{%
 \ifx #1\expandafter \@firstoftwo
 \else \expandafter \@secondoftwo
 \fi
}%
\providecommand \natexlab [1]{#1}%
\providecommand \enquote  [1]{``#1''}%
\providecommand \bibnamefont  [1]{#1}%
\providecommand \bibfnamefont [1]{#1}%
\providecommand \citenamefont [1]{#1}%
\providecommand \href@noop [0]{\@secondoftwo}%
\providecommand \href [0]{\begingroup \@sanitize@url \@href}%
\providecommand \@href[1]{\@@startlink{#1}\@@href}%
\providecommand \@@href[1]{\endgroup#1\@@endlink}%
\providecommand \@sanitize@url [0]{\catcode `\\12\catcode `\$12\catcode
  `\&12\catcode `\#12\catcode `\^12\catcode `\_12\catcode `\%12\relax}%
\providecommand \@@startlink[1]{}%
\providecommand \@@endlink[0]{}%
\providecommand \url  [0]{\begingroup\@sanitize@url \@url }%
\providecommand \@url [1]{\endgroup\@href {#1}{\urlprefix }}%
\providecommand \urlprefix  [0]{URL }%
\providecommand \Eprint [0]{\href }%
\providecommand \doibase [0]{http://dx.doi.org/}%
\providecommand \selectlanguage [0]{\@gobble}%
\providecommand \bibinfo  [0]{\@secondoftwo}%
\providecommand \bibfield  [0]{\@secondoftwo}%
\providecommand \translation [1]{[#1]}%
\providecommand \BibitemOpen [0]{}%
\providecommand \bibitemStop [0]{}%
\providecommand \bibitemNoStop [0]{.\EOS\space}%
\providecommand \EOS [0]{\spacefactor3000\relax}%
\providecommand \BibitemShut  [1]{\csname bibitem#1\endcsname}%
\let\auto@bib@innerbib\@empty
\bibitem [{\citenamefont {Katsnelson}\ and\ \citenamefont
  {Kat?s?nel?son}(2012)}]{r1}%
  \BibitemOpen
  \bibfield  {author} {\bibinfo {author} {\bibfnamefont {M.~I.}\ \bibnamefont
  {Katsnelson}}\ and\ \bibinfo {author} {\bibfnamefont {M.~I.}\ \bibnamefont
  {Kat?s?nel?son}},\ }\href@noop {} {\emph {\bibinfo {title} {Graphene: carbon
  in two dimensions}}}\ (\bibinfo  {publisher} {Cambridge University Press},\
  \bibinfo {year} {2012})\BibitemShut {NoStop}%
\bibitem [{\citenamefont {Goerbig}(2011)}]{r2}%
  \BibitemOpen
  \bibfield  {author} {\bibinfo {author} {\bibfnamefont {M.~O.}\ \bibnamefont
  {Goerbig}},\ }\href@noop {} {\bibfield  {journal} {\bibinfo  {journal}
  {Reviews of Modern Physics}\ }\textbf {\bibinfo {volume} {83}},\ \bibinfo
  {pages} {1193} (\bibinfo {year} {2011})}\BibitemShut {NoStop}%
\bibitem [{\citenamefont {Falkovsky}(2012)}]{r3}%
  \BibitemOpen
  \bibfield  {author} {\bibinfo {author} {\bibfnamefont {L.}~\bibnamefont
  {Falkovsky}},\ }\href@noop {} {\bibfield  {journal} {\bibinfo  {journal}
  {Journal of Experimental and Theoretical Physics}\ }\textbf {\bibinfo
  {volume} {115}},\ \bibinfo {pages} {1151} (\bibinfo {year}
  {2012})}\BibitemShut {NoStop}%
\bibitem [{\citenamefont {Falkovsky}(2013)}]{r4}%
  \BibitemOpen
  \bibfield  {author} {\bibinfo {author} {\bibfnamefont {L.~A.}\ \bibnamefont
  {Falkovsky}},\ }\href@noop {} {\bibfield  {journal} {\bibinfo  {journal}
  {JETP letters}\ }\textbf {\bibinfo {volume} {97}},\ \bibinfo {pages} {429}
  (\bibinfo {year} {2013})}\BibitemShut {NoStop}%
\bibitem [{\citenamefont {LukÕyanchuk}\ \emph
  {et~al.}(2011{\natexlab{a}})\citenamefont {LukÕyanchuk}, \citenamefont
  {Varlamov},\ and\ \citenamefont {Kavokin}}]{r5}%
  \BibitemOpen
  \bibfield  {author} {\bibinfo {author} {\bibfnamefont {I.~A.}\ \bibnamefont
  {LukÕyanchuk}}, \bibinfo {author} {\bibfnamefont {A.~A.}\ \bibnamefont
  {Varlamov}}, \ and\ \bibinfo {author} {\bibfnamefont {A.~V.}\ \bibnamefont
  {Kavokin}},\ }\href@noop {} {\bibfield  {journal} {\bibinfo  {journal}
  {Physical review letters}\ }\textbf {\bibinfo {volume} {107}},\ \bibinfo
  {pages} {016601} (\bibinfo {year} {2011}{\natexlab{a}})}\BibitemShut
  {NoStop}%
\bibitem [{\citenamefont {Novoselov}\ \emph {et~al.}(2005)\citenamefont
  {Novoselov}, \citenamefont {Geim}, \citenamefont {Morozov}, \citenamefont
  {Jiang}, \citenamefont {Katsnelson}, \citenamefont {Grigorieva},
  \citenamefont {Dubonos},\ and\ \citenamefont {Firsov}}]{r6}%
  \BibitemOpen
  \bibfield  {author} {\bibinfo {author} {\bibfnamefont {K.}~\bibnamefont
  {Novoselov}}, \bibinfo {author} {\bibfnamefont {A.~K.}\ \bibnamefont {Geim}},
  \bibinfo {author} {\bibfnamefont {S.}~\bibnamefont {Morozov}}, \bibinfo
  {author} {\bibfnamefont {D.}~\bibnamefont {Jiang}}, \bibinfo {author}
  {\bibfnamefont {M.}~\bibnamefont {Katsnelson}}, \bibinfo {author}
  {\bibfnamefont {I.}~\bibnamefont {Grigorieva}}, \bibinfo {author}
  {\bibfnamefont {S.}~\bibnamefont {Dubonos}}, \ and\ \bibinfo {author}
  {\bibfnamefont {A.}~\bibnamefont {Firsov}},\ }\href@noop {} {\bibfield
  {journal} {\bibinfo  {journal} {nature}\ }\textbf {\bibinfo {volume} {438}},\
  \bibinfo {pages} {197} (\bibinfo {year} {2005})}\BibitemShut {NoStop}%
\bibitem [{\citenamefont {Gusynin}\ and\ \citenamefont {Sharapov}(2005)}]{r7}%
  \BibitemOpen
  \bibfield  {author} {\bibinfo {author} {\bibfnamefont {V.}~\bibnamefont
  {Gusynin}}\ and\ \bibinfo {author} {\bibfnamefont {S.}~\bibnamefont
  {Sharapov}},\ }\href@noop {} {\bibfield  {journal} {\bibinfo  {journal}
  {Physical Review Letters}\ }\textbf {\bibinfo {volume} {95}},\ \bibinfo
  {pages} {146801} (\bibinfo {year} {2005})}\BibitemShut {NoStop}%
\bibitem [{\citenamefont {LukÕyanchuk}\ \emph
  {et~al.}(2011{\natexlab{b}})\citenamefont {LukÕyanchuk}, \citenamefont
  {Varlamov},\ and\ \citenamefont {Kavokin}}]{cinco}%
  \BibitemOpen
  \bibfield  {author} {\bibinfo {author} {\bibfnamefont {I.~A.}\ \bibnamefont
  {LukÕyanchuk}}, \bibinfo {author} {\bibfnamefont {A.~A.}\ \bibnamefont
  {Varlamov}}, \ and\ \bibinfo {author} {\bibfnamefont {A.~V.}\ \bibnamefont
  {Kavokin}},\ }\href@noop {} {\bibfield  {journal} {\bibinfo  {journal}
  {Physical review letters}\ }\textbf {\bibinfo {volume} {107}},\ \bibinfo
  {pages} {016601} (\bibinfo {year} {2011}{\natexlab{b}})}\BibitemShut
  {NoStop}%
\bibitem [{\citenamefont {Alisultanov}\ and\ \citenamefont
  {Reis}(2015{\natexlab{a}})}]{alisultanov2015oscillating}%
  \BibitemOpen
  \bibfield  {author} {\bibinfo {author} {\bibfnamefont {Z.}~\bibnamefont
  {Alisultanov}}\ and\ \bibinfo {author} {\bibfnamefont {M.}~\bibnamefont
  {Reis}},\ }\href@noop {} {\bibfield  {journal} {\bibinfo  {journal} {Solid
  State Communications}\ }\textbf {\bibinfo {volume} {206}},\ \bibinfo {pages}
  {17} (\bibinfo {year} {2015}{\natexlab{a}})}\BibitemShut {NoStop}%
\bibitem [{\citenamefont {Alisultanov}\ \emph {et~al.}(2014)\citenamefont
  {Alisultanov}, \citenamefont {Paixao},\ and\ \citenamefont
  {Reis}}]{alisultanov2014oscillating}%
  \BibitemOpen
  \bibfield  {author} {\bibinfo {author} {\bibfnamefont {Z.}~\bibnamefont
  {Alisultanov}}, \bibinfo {author} {\bibfnamefont {L.}~\bibnamefont {Paixao}},
  \ and\ \bibinfo {author} {\bibfnamefont {M.}~\bibnamefont {Reis}},\
  }\href@noop {} {\bibfield  {journal} {\bibinfo  {journal} {Applied Physics
  Letters}\ }\textbf {\bibinfo {volume} {105}},\ \bibinfo {pages} {232406}
  (\bibinfo {year} {2014})}\BibitemShut {NoStop}%
\bibitem [{\citenamefont {Paix{\~a}o}\ \emph {et~al.}(2014)\citenamefont
  {Paix{\~a}o}, \citenamefont {Alisultanov},\ and\ \citenamefont
  {Reis}}]{paixao2014oscillating}%
  \BibitemOpen
  \bibfield  {author} {\bibinfo {author} {\bibfnamefont {L.}~\bibnamefont
  {Paix{\~a}o}}, \bibinfo {author} {\bibfnamefont {Z.}~\bibnamefont
  {Alisultanov}}, \ and\ \bibinfo {author} {\bibfnamefont {M.}~\bibnamefont
  {Reis}},\ }\href@noop {} {\bibfield  {journal} {\bibinfo  {journal} {Journal
  of Magnetism and Magnetic Materials}\ }\textbf {\bibinfo {volume} {368}},\
  \bibinfo {pages} {374} (\bibinfo {year} {2014})}\BibitemShut {NoStop}%
\bibitem [{\citenamefont {Reis}(2014{\natexlab{a}})}]{reis2014diamagnetic}%
  \BibitemOpen
  \bibfield  {author} {\bibinfo {author} {\bibfnamefont {M.}~\bibnamefont
  {Reis}},\ }\href@noop {} {\bibfield  {journal} {\bibinfo  {journal} {Physics
  Letters A}\ }\textbf {\bibinfo {volume} {378}},\ \bibinfo {pages} {1903}
  (\bibinfo {year} {2014}{\natexlab{a}})}\BibitemShut {NoStop}%
\bibitem [{\citenamefont {Reis}(2014{\natexlab{b}})}]{reis2014step}%
  \BibitemOpen
  \bibfield  {author} {\bibinfo {author} {\bibfnamefont {M.}~\bibnamefont
  {Reis}},\ }\href@noop {} {\bibfield  {journal} {\bibinfo  {journal} {Physics
  Letters A}\ }\textbf {\bibinfo {volume} {378}},\ \bibinfo {pages} {918}
  (\bibinfo {year} {2014}{\natexlab{b}})}\BibitemShut {NoStop}%
\bibitem [{\citenamefont {Reis}(2013{\natexlab{a}})}]{reis2013oscillating}%
  \BibitemOpen
  \bibfield  {author} {\bibinfo {author} {\bibfnamefont {M.}~\bibnamefont
  {Reis}},\ }\href@noop {} {\bibfield  {journal} {\bibinfo  {journal} {Journal
  of Applied Physics}\ }\textbf {\bibinfo {volume} {113}},\ \bibinfo {pages}
  {243901} (\bibinfo {year} {2013}{\natexlab{a}})}\BibitemShut {NoStop}%
\bibitem [{\citenamefont {Reis}\ and\ \citenamefont
  {Soriano}(2013)}]{reis2013electrocaloric}%
  \BibitemOpen
  \bibfield  {author} {\bibinfo {author} {\bibfnamefont {M.}~\bibnamefont
  {Reis}}\ and\ \bibinfo {author} {\bibfnamefont {S.}~\bibnamefont {Soriano}},\
  }\href@noop {} {\bibfield  {journal} {\bibinfo  {journal} {Applied Physics
  Letters}\ }\textbf {\bibinfo {volume} {102}},\ \bibinfo {pages} {112903}
  (\bibinfo {year} {2013})}\BibitemShut {NoStop}%
\bibitem [{\citenamefont {Reis}(2013{\natexlab{b}})}]{reis2013influence}%
  \BibitemOpen
  \bibfield  {author} {\bibinfo {author} {\bibfnamefont {M.}~\bibnamefont
  {Reis}},\ }\href@noop {} {\bibfield  {journal} {\bibinfo  {journal} {Solid
  State Communications}\ }\textbf {\bibinfo {volume} {161}},\ \bibinfo {pages}
  {19} (\bibinfo {year} {2013}{\natexlab{b}})}\BibitemShut {NoStop}%
\bibitem [{\citenamefont {Reis}(2012)}]{reis2012oscillating}%
  \BibitemOpen
  \bibfield  {author} {\bibinfo {author} {\bibfnamefont {M.}~\bibnamefont
  {Reis}},\ }\href@noop {} {\bibfield  {journal} {\bibinfo  {journal} {Solid
  State Communications}\ }\textbf {\bibinfo {volume} {152}},\ \bibinfo {pages}
  {921} (\bibinfo {year} {2012})}\BibitemShut {NoStop}%
\bibitem [{\citenamefont {Reis}(2011)}]{reis2011oscillating}%
  \BibitemOpen
  \bibfield  {author} {\bibinfo {author} {\bibfnamefont {M.}~\bibnamefont
  {Reis}},\ }\href@noop {} {\bibfield  {journal} {\bibinfo  {journal} {Applied
  Physics Letters}\ }\textbf {\bibinfo {volume} {99}},\ \bibinfo {pages}
  {052511} (\bibinfo {year} {2011})}\BibitemShut {NoStop}%
\bibitem [{\citenamefont {Reis}(2015)}]{reis2015magnetocaloric}%
  \BibitemOpen
  \bibfield  {author} {\bibinfo {author} {\bibfnamefont {M.}~\bibnamefont
  {Reis}},\ }\href@noop {} {\bibfield  {journal} {\bibinfo  {journal} {Applied
  Physics Letters}\ }\textbf {\bibinfo {volume} {107}},\ \bibinfo {pages}
  {102401} (\bibinfo {year} {2015})}\BibitemShut {NoStop}%
\bibitem [{\citenamefont {Alisultanov}(2014{\natexlab{a}})}]{r16}%
  \BibitemOpen
  \bibfield  {author} {\bibinfo {author} {\bibfnamefont {Z.~Z.}\ \bibnamefont
  {Alisultanov}},\ }\href@noop {} {\bibfield  {journal} {\bibinfo  {journal}
  {JETP letters}\ }\textbf {\bibinfo {volume} {99}},\ \bibinfo {pages} {232}
  (\bibinfo {year} {2014}{\natexlab{a}})}\BibitemShut {NoStop}%
\bibitem [{\citenamefont {Lifshits}\ and\ \citenamefont {Koganov}(1959)}]{r15}%
  \BibitemOpen
  \bibfield  {author} {\bibinfo {author} {\bibfnamefont {I.}~\bibnamefont
  {Lifshits}}\ and\ \bibinfo {author} {\bibfnamefont {M.}~\bibnamefont
  {Koganov}},\ }\href@noop {} {\bibfield  {journal} {\bibinfo  {journal}
  {Uspekhi Fiz. Nauk}\ }\textbf {\bibinfo {volume} {69}} (\bibinfo {year}
  {1959})}\BibitemShut {NoStop}%
\bibitem [{\citenamefont {Alisultanov}(2014{\natexlab{b}})}]{r18}%
  \BibitemOpen
  \bibfield  {author} {\bibinfo {author} {\bibfnamefont {Z.~Z.}\ \bibnamefont
  {Alisultanov}},\ }\href@noop {} {\bibfield  {journal} {\bibinfo  {journal}
  {JETP letters}\ }\textbf {\bibinfo {volume} {99}},\ \bibinfo {pages} {702}
  (\bibinfo {year} {2014}{\natexlab{b}})}\BibitemShut {NoStop}%
\bibitem [{\citenamefont {Lifshitz}\ and\ \citenamefont
  {Kosevich}(1955)}]{r19}%
  \BibitemOpen
  \bibfield  {author} {\bibinfo {author} {\bibfnamefont {I.}~\bibnamefont
  {Lifshitz}}\ and\ \bibinfo {author} {\bibfnamefont {A.}~\bibnamefont
  {Kosevich}},\ }\href@noop {} {\bibfield  {journal} {\bibinfo  {journal}
  {English translation 1956 Soviet Phys. JETP}\ }\textbf {\bibinfo {volume}
  {2}},\ \bibinfo {pages} {636} (\bibinfo {year} {1955})}\BibitemShut {NoStop}%
\bibitem [{\citenamefont {Onsager}(1952)}]{r20}%
  \BibitemOpen
  \bibfield  {author} {\bibinfo {author} {\bibfnamefont {L.}~\bibnamefont
  {Onsager}},\ }\href@noop {} {\bibfield  {journal} {\bibinfo  {journal} {The
  London, Edinburgh, and Dublin Philosophical Magazine and Journal of Science}\
  }\textbf {\bibinfo {volume} {43}},\ \bibinfo {pages} {1006} (\bibinfo {year}
  {1952})}\BibitemShut {NoStop}%
\bibitem [{\citenamefont {Alisultanov}\ and\ \citenamefont
  {Reis}(2015{\natexlab{b}})}]{alisultanov2015magneto}%
  \BibitemOpen
  \bibfield  {author} {\bibinfo {author} {\bibfnamefont {Z.}~\bibnamefont
  {Alisultanov}}\ and\ \bibinfo {author} {\bibfnamefont {M.}~\bibnamefont
  {Reis}},\ }\href@noop {} {\bibfield  {journal} {\bibinfo  {journal} {arXiv
  preprint arXiv:1508.06648}\ } (\bibinfo {year}
  {2015}{\natexlab{b}})}\BibitemShut {NoStop}%
\bibitem [{\citenamefont {Lukose}\ \emph {et~al.}(2007)\citenamefont {Lukose},
  \citenamefont {Shankar},\ and\ \citenamefont {Baskaran}}]{r13}%
  \BibitemOpen
  \bibfield  {author} {\bibinfo {author} {\bibfnamefont {V.}~\bibnamefont
  {Lukose}}, \bibinfo {author} {\bibfnamefont {R.}~\bibnamefont {Shankar}}, \
  and\ \bibinfo {author} {\bibfnamefont {G.}~\bibnamefont {Baskaran}},\
  }\href@noop {} {\bibfield  {journal} {\bibinfo  {journal} {Physical review
  letters}\ }\textbf {\bibinfo {volume} {98}},\ \bibinfo {pages} {116802}
  (\bibinfo {year} {2007})}\BibitemShut {NoStop}%
\bibitem [{\citenamefont {Gusynin}\ \emph {et~al.}(2014)\citenamefont
  {Gusynin}, \citenamefont {Loktev}, \citenamefont {Luk'yanchuk}, \citenamefont
  {Sharapov},\ and\ \citenamefont {Varlamov}}]{r10}%
  \BibitemOpen
  \bibfield  {author} {\bibinfo {author} {\bibfnamefont {V.}~\bibnamefont
  {Gusynin}}, \bibinfo {author} {\bibfnamefont {V.}~\bibnamefont {Loktev}},
  \bibinfo {author} {\bibfnamefont {I.}~\bibnamefont {Luk'yanchuk}}, \bibinfo
  {author} {\bibfnamefont {S.}~\bibnamefont {Sharapov}}, \ and\ \bibinfo
  {author} {\bibfnamefont {A.}~\bibnamefont {Varlamov}},\ }\href@noop {}
  {\bibfield  {journal} {\bibinfo  {journal} {Low Temperature Physics}\
  }\textbf {\bibinfo {volume} {40}},\ \bibinfo {pages} {270} (\bibinfo {year}
  {2014})}\BibitemShut {NoStop}%
\bibitem [{\citenamefont {Asgari}\ \emph {et~al.}(2014)\citenamefont {Asgari},
  \citenamefont {Katsnelson},\ and\ \citenamefont {Polini}}]{r11}%
  \BibitemOpen
  \bibfield  {author} {\bibinfo {author} {\bibfnamefont {R.}~\bibnamefont
  {Asgari}}, \bibinfo {author} {\bibfnamefont {M.~I.}\ \bibnamefont
  {Katsnelson}}, \ and\ \bibinfo {author} {\bibfnamefont {M.}~\bibnamefont
  {Polini}},\ }\href@noop {} {\bibfield  {journal} {\bibinfo  {journal}
  {Annalen der Physik}\ }\textbf {\bibinfo {volume} {526}},\ \bibinfo {pages}
  {359} (\bibinfo {year} {2014})}\BibitemShut {NoStop}%
\bibitem [{\citenamefont {Lozovik}\ \emph {et~al.}(2015)\citenamefont
  {Lozovik}, \citenamefont {Sokolik},\ and\ \citenamefont {Zabolotskiy}}]{r12}%
  \BibitemOpen
  \bibfield  {author} {\bibinfo {author} {\bibfnamefont {Y.~E.}\ \bibnamefont
  {Lozovik}}, \bibinfo {author} {\bibfnamefont {A.}~\bibnamefont {Sokolik}}, \
  and\ \bibinfo {author} {\bibfnamefont {A.}~\bibnamefont {Zabolotskiy}},\
  }\href@noop {} {\bibfield  {journal} {\bibinfo  {journal} {Physical Review
  B}\ }\textbf {\bibinfo {volume} {91}},\ \bibinfo {pages} {075416} (\bibinfo
  {year} {2015})}\BibitemShut {NoStop}%
\bibitem [{\citenamefont {Aronov}\ and\ \citenamefont
  {Pikus}(1966)}]{aronov1966indirect}%
  \BibitemOpen
  \bibfield  {author} {\bibinfo {author} {\bibfnamefont {A.}~\bibnamefont
  {Aronov}}\ and\ \bibinfo {author} {\bibfnamefont {G.}~\bibnamefont {Pikus}},\
  }\href@noop {} {\bibfield  {journal} {\bibinfo  {journal} {SOVIET PHYSICS
  JETP}\ }\textbf {\bibinfo {volume} {22}} (\bibinfo {year}
  {1966})}\BibitemShut {NoStop}%
\end{thebibliography}
\end{document}